\documentclass{nature}

\usepackage{color}
\usepackage{amssymb}
\usepackage{graphicx}
\usepackage{hyperref}
\usepackage{bibunits}
\defaultbibliography{KSN2015K}
\defaultbibliographystyle{naturemag}

\newif\ifsuppl
\supplfalse
\suppltrue                                                                     

\newif\ifmaintext
\maintextfalse
\maintexttrue

\def\reff@jnl#1{{\rm#1\/}}
\def\apj{\reff@jnl{ApJ}}
\def\aj{\reff@jnl{AJ}}                  
\def\araa{\reff@jnl{ARA\&A}}            
\def\apj{\reff@jnl{ApJ}}                
\def\apjl{\reff@jnl{ApJ}}               
\def\apjs{\reff@jnl{ApJS}}              
\def\aap{\reff@jnl{A\&A}}               
\def\mnras{\reff@jnl{MNRAS}}            
\def\prd{\reff@jnl{Phys.Rev.D}}         
\def\prl{\reff@jnl{Phys.Rev.Lett}}      
\def\pasp{\reff@jnl{PASP}}              
\def\nat{\reff@jnl{Nature}}             
\def\iaucirc{\reff@jnl{IAU~Circ.}}%
\def\aaps{\reff@jnl{A\&AS}}%
\def\pasa{\reff@jnl{PASA}}%
\newcommand\actaa{\ref@jnl{Acta Astron.}}%

\newcommand{\Msun}[0]{
\ensuremath{M_{\odot}}}

\title{A Fast-Evolving, Luminous Transient Discovered by K2/Kepler}

\author
{A. Rest,$^{1,2}$ 
P. M. Garnavich,$^3$
D. Khatami,$^{4,5}$ 
D. Kasen,$^{4,5}$ 
B. E. Tucker,$^{6,7}$
E. J. Shaya,$^8$
R. P. Olling,$^8$ 
R. Mushotzky,$^8$
A. Zenteno,$^9$
S. Margheim,$^10$
G. Strampelli,$^1$ 
D. James,$^{11}$
R. C. Smith,$^9$
F. F\"orster,$^{12}$
V. A. Villar$^{11}$
}

\begin{document}

\maketitle

\begin{affiliations}
\item Space Telescope Science Institute, 3700 San Martin Drive, Baltimore, MD 21218, USA.
\item Department of Physics and Astronomy, The Johns Hopkins University, 3400 North Charles Street, Baltimore, MD 21218, USA.
\item Department of Physics, University of Notre Dame, 225 Nieuwland Science Hall, Notre Dame, IN, 46556-5670, USA.
 \item Department of Astronomy, University of California, Berkeley, CA 94720-3411, USA.
\item Lawrence Berkeley National Laboratory, 1 Cyclotron Road, Berkeley, California 94720, USA.
\item The Research School of Astronomy and Astrophysics, Mount Stromlo Observatory, Australian National University, via Cotter Road, Canberra, ACT 2611, Australia.
\item The ARC Centre of Excellence for All-Sky Astrophysics (CAASTRO)
\item Astronomy Department, University of Maryland, College Park, MD 20742-2421, USA.
\item Cerro Tololo Inter-American Observatory, Casilla 603, La Serena, Chile.
\item Gemini Observatory, La Serena, Chile.
\item Harvard-Smithsonian Center for Astrophysics, 60 Garden Street, Cambridge, MA 02138, USA.
\item Center for Mathematical Modeling, University of Chile, Santiago, Chile.
\end{affiliations}

\begin{bibunit}
\begin{abstract}
For decades optical time-domain searches have been tuned to find ordinary supernovae, which rise and fall in brightness over a period of weeks. Recently, supernova searches have improved their cadences and a handful of fast-evolving luminous transients (FELTs) have been identified\cite{poznanski10,kasliwal10,drout14,shivvers16,arcavi16}. FELTs have peak luminosities comparable to Type~Ia supernovae, but rise to maximum in $<10$~days and fade from view in $<$month. Here we present the most extreme example of this class thus far, KSN2015K, with a rise time of only 2.2~days and a time above half-maximum ($t_{1/2}$) of only 6.8~days. Here we show that, unlike Type~Ia supernovae, the light curve of KSN2015K was not powered by the decay of radioactive elements. We further argue that it is unlikely that it was powered by continuing energy deposition from a central remnant (a magnetar or black hole). Using numerical radiation hydrodynamical models, we show that the light curve of KSN2015K is well fit by a model where the supernova runs into external material presumably expelled in a pre-supernova mass loss episode. The rapid rise of KSN2015K therefore probes the venting of photons when a hypersonic shock wave breaks out of a dense extended medium.
\end{abstract}

We identified KSN2015K as an unusual transient in the K2 Campaign~6 data from the extended Kepler mission\cite{howell14}. While we have several ground-based optical programs to find supernovae during a K2 Campaign, KSN2015K was identified in February 2016 after the Campaign~6 data were publicly released. Re-analysis of images taken by the Dark Energy Camera (DECam) and SkyMapper clearly show the transient, but it was not flagged because it only appeared on one epoch. We therefore could not obtain a spectrum of the transient itself. The host has a redshift of 0.090 implying a luminosity distance of 410~Mpc (assuming a flat cosmology with H$_0=70$~km~s$^{-1}$~Mpc$^{-1}$).  

\begin{figure}
\includegraphics[scale=0.65]{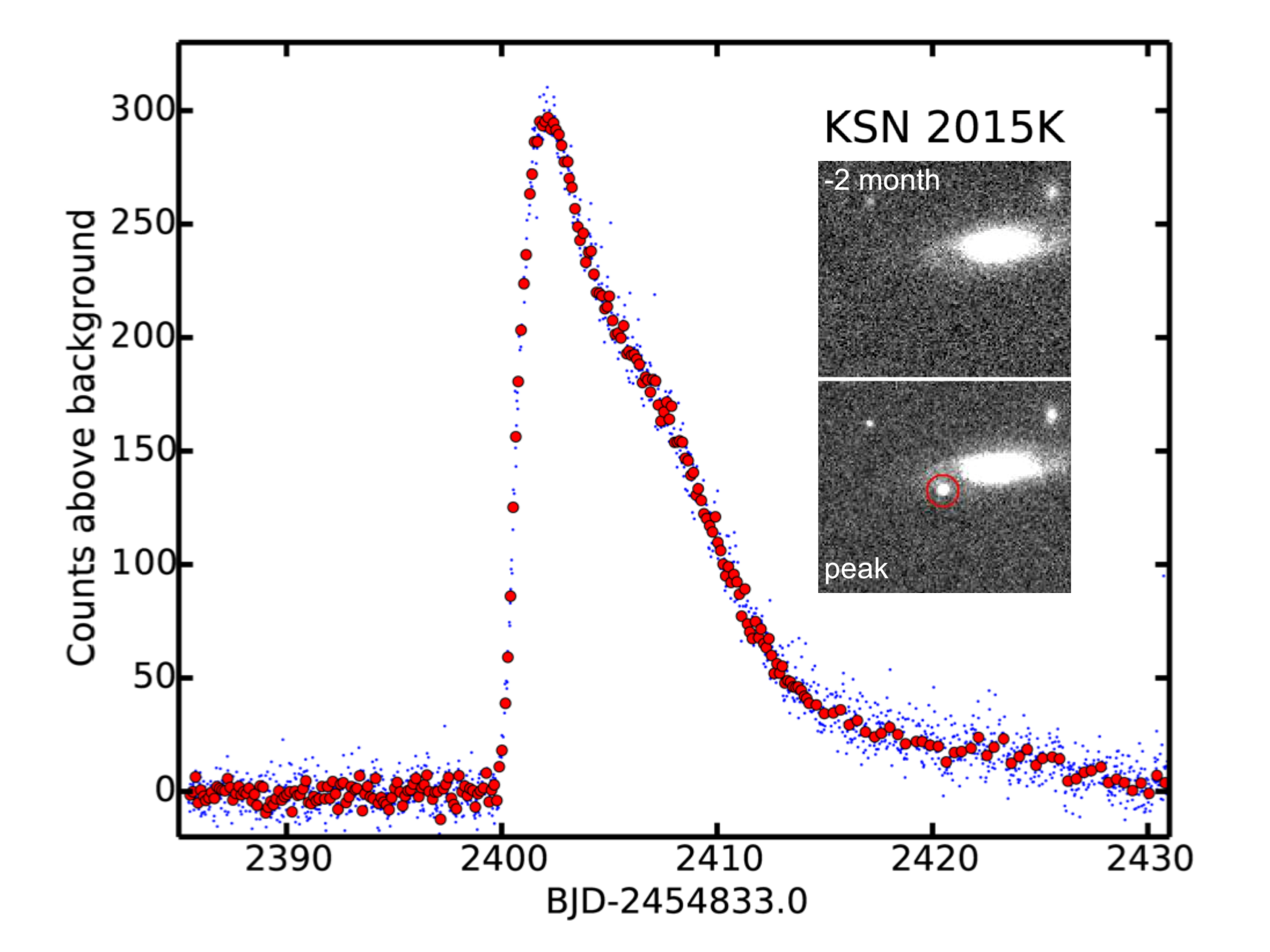}
\caption{{\bf The K2 light curve of KSN2015K.} Blue dots are individual 30-minute cadence observations while the red points represent 3-hour median-value bins. The image cutouts in the inset show 60 second $i$-band DECam images from UT July 7th 2015 (2 months before peak) and August 1st 2015 (around peak) in the top and bottom panels, respectively. KSN2015K is marked with a red circle in the bottom panel. The photometric uncertainty is seen as the scatter of the K2 observations before the outburst.}
\label{fig:lightcurve}
\end{figure}

\begin{figure}
\includegraphics[scale=0.65]{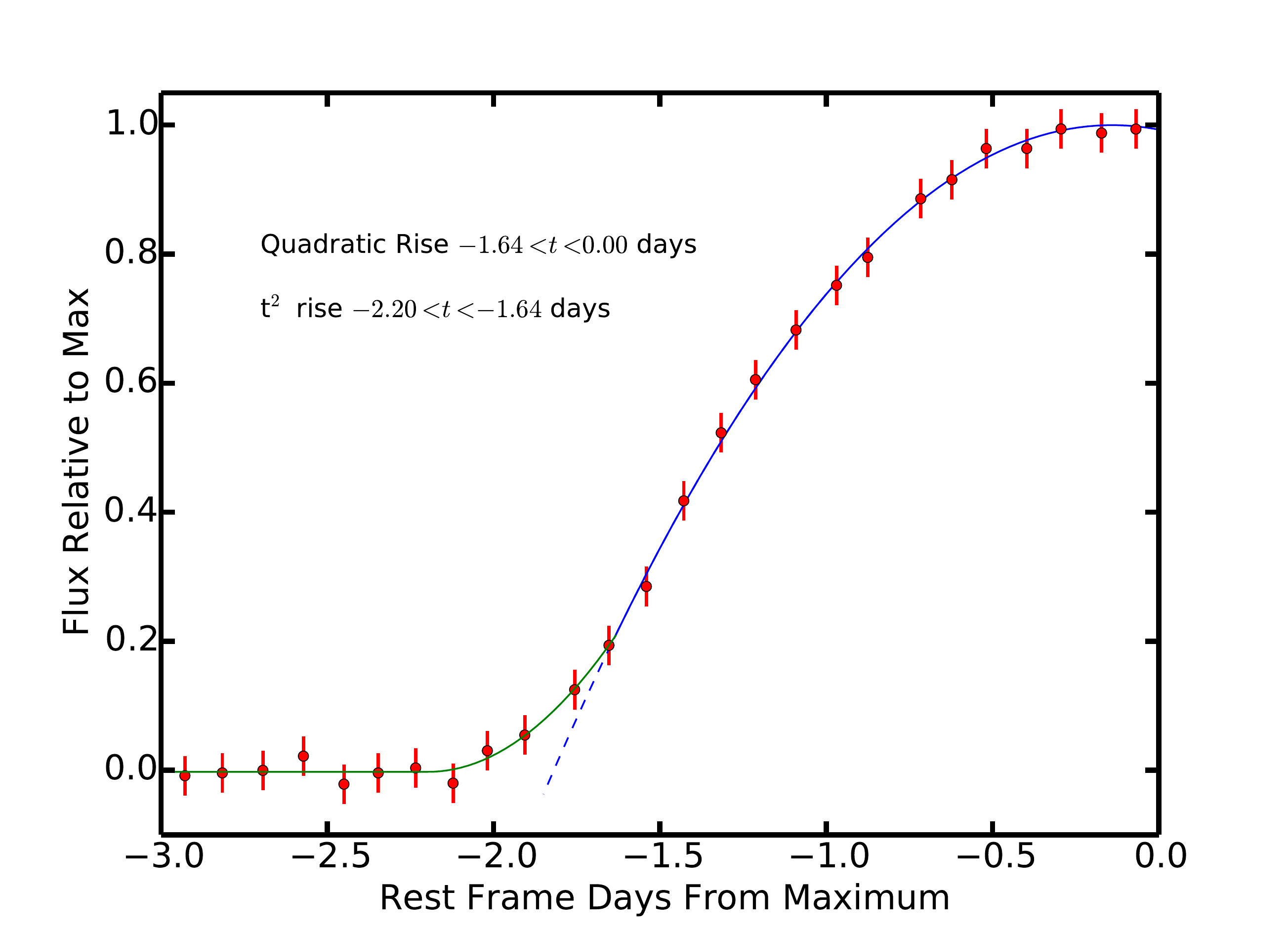}
\caption{{\bf KSN2015K's rise to maximum light.} Red points are 3 hour median bins of the K2 long cadence data. Error bars are $3\sigma$ uncertainties on the binned photometry points. The blue line is a quadratic fit to the points between $-1.64<t<0.0$ days. The green line is a $(t-t_0)^2$ fit to data between $-2.2<t<-1.64$ days. The uncertainty on each point is estimated from the scatter of the six measurements averaged in each bin.  } 
\label{fig:rise}
\end{figure}

The K2 light curve of KSN2015K seems to have four phases (see Figures~\ref{fig:lightcurve}~and~\ref{fig:rise}). The rise is well fit by a quadratic function starting 1.6 days before maximum. Before that, the light rises like $t^2$ and suggest the explosion occurred $2.2\pm 0.1$ days before peak brightness. After maximum, KSN2015K shows a decline followed by a plateau and finally a power-law decay. Additional, ground-based photometry from DECam and SkyMapper show the color to be quite blue (see supplemental material). At peak, KSN2015K's color is $r-i=-0.15\pm0.05$, and $\sim8$ days after peak its color remains quite blue at $g-r=-0.17\pm0.20$ even after fading to half its peak brightness.

KSN2015K's host is a star-forming spiral galaxy and the transient is seen projected on a spiral arm (see Figure~\ref{fig:lightcurve} and the Methods section). If the transient is associated with the arm, the environment suggests a relatively short time between birth and the transient outburst, but both thermonuclear and core-collapse supernovae are found in young, star-forming populations.

The progenitors of FELTs and the energy source that powers the light curve have been debated. Members of the class could originate from more than one type of progenitor. The high-cadence time-sampling of KSN2015K allows us to establish strong constraints on the origin of this particular event.

If FELT light curves are powered by an internal energy source, such as the decay of radioactive isotopes or a central engine, then the light curve rise time is set by the photon diffusion timescale through the remnant, $t_{\rm diff}\propto(M\kappa/v)^{1/2}$ where $M$ is the remnant mass, $v$ the expansion velocity, and $\kappa$ the opacity. The rise time of KSN2015k was $\sim8$ times shorter than that of Type~Ia supernovae (which have mass $\approx1~\Msun$), implying an ejected mass of only a few times $\sim10^{-2}~M_\odot$. Differences in the velocity and opacity of the ejecta are unlikely to change this estimate by more than a factor of several. This mass constraint, however, does not apply to mechanisms that directly deposit thermal energy near the ejecta surface, such as the blast wave from the explosion or shocks from circumstellar interaction.

\begin{figure}
\includegraphics[scale=0.65]{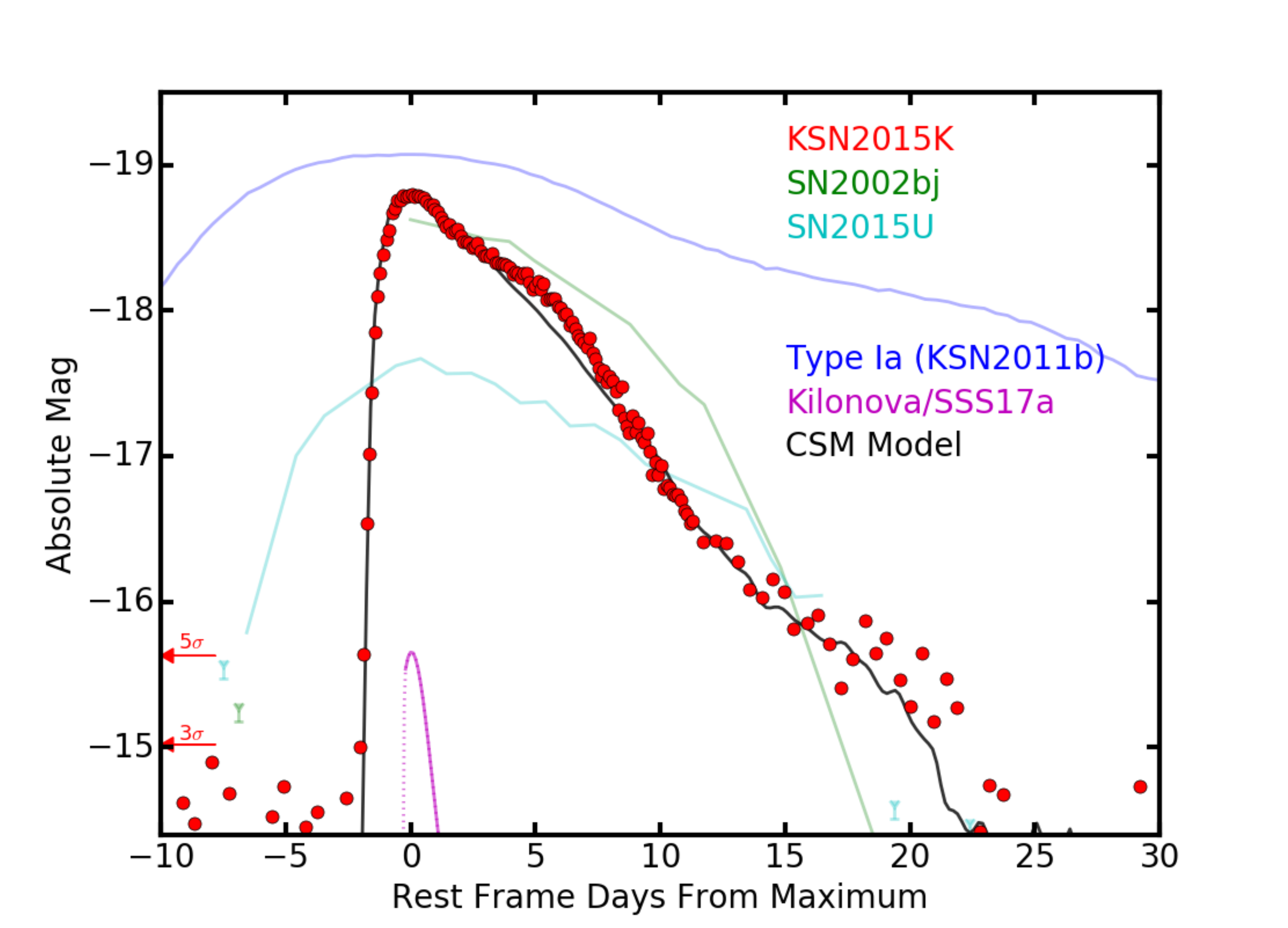}
\caption{{\bf Light curve comparison.} The KSN2015K light curve (red dots) assuming H$_0=70$ km~s$^{-1}$~Mpc$^{-1}$ and a Milky Way extinction of A$_V=0.10$ mag. The light curve of another Kepler type~Ia supernova (blue line) is shown for comparison. Also shown are light curves of the fast transients SN2002bj and SN2015U, and the kilonova AT2017gfo/SSS17a\cite{Abbott17,Villar17}. The black line shows the best fit shock breakout in circumstellar material model. The detection significance for KSN2015K is indicated at the lower left of the figure as the number of standard deviations ($\sigma$) from the average background.}
\label{fig:absmag}
\end{figure}

\begin{figure}
\includegraphics[scale=0.60]{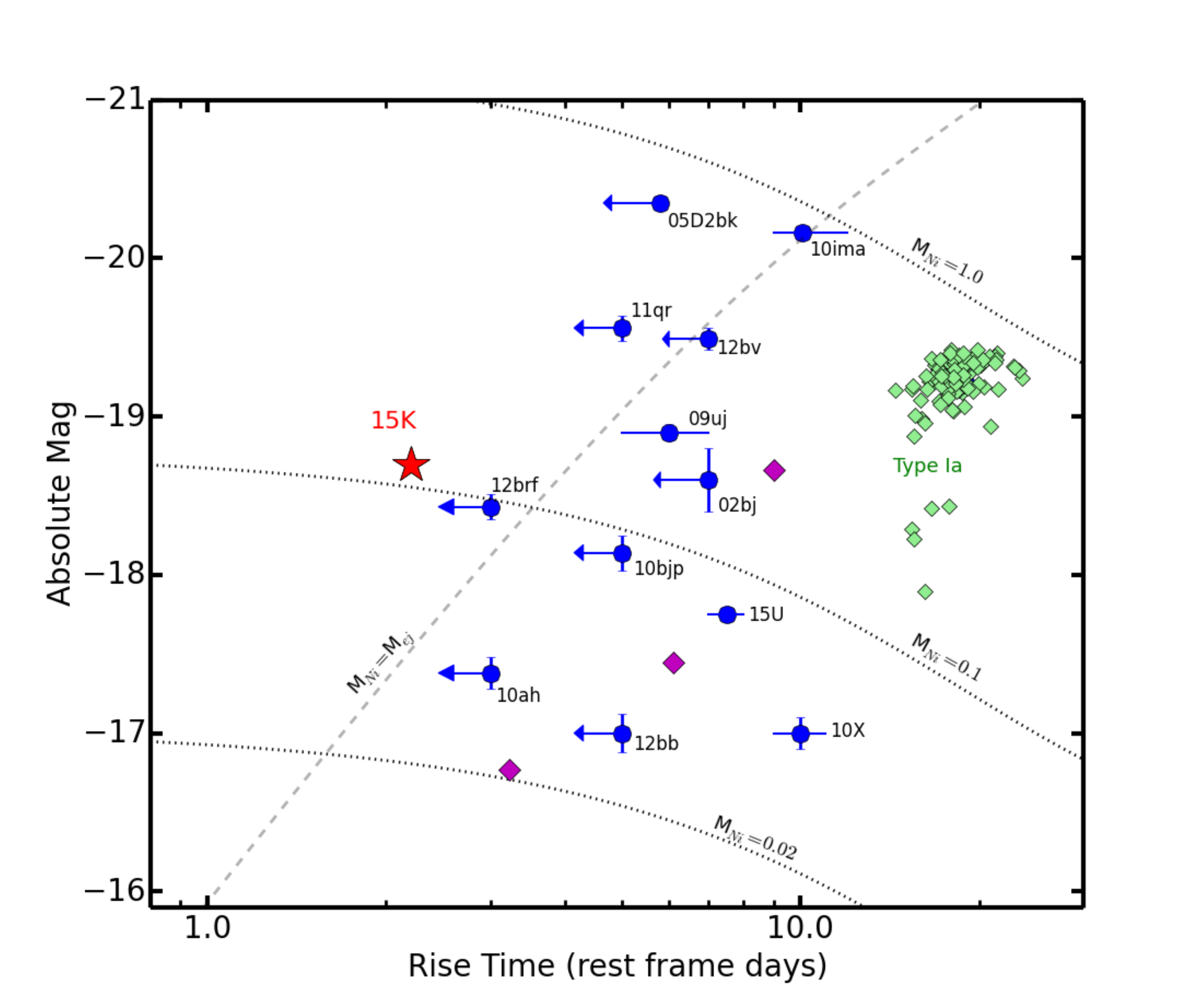}
\caption{{\bf Peak luminosity versus rise time:} The peak luminosity versus rise time at optical wavelengths for fast transients (blue) and type Ia supernovae (green) from SDSS-II. The red star shows the position of KSN2015K. Purple diamonds show ``.1a'' models\cite{shen10}. Dotted lines show Arnett's rule for a range of synthesized $^{56}$Ni masses. The dashed line is a thermonuclear scenario where a pure $^{56}$Ni envelope is ejected at 10000~km~s$^{-1}$. Events to the left of the dashed line cannot be fully powered with radioactive decay. The errors on the rise times are taken from the literature. For KSN2015K, the uncertainty is smaller than the symbol size and is estimated from the data show in Figure~\ref{fig:rise}.
} 
\label{fig:risetime}
\end{figure}

There are several explosive scenarios that may lead to the ejection of such a small radioactive mass ($10^{-4}-10^{-1}$~\Msun), such as the thermonuclear explosion  of a shell of accreted Helium on the surface of a white dwarf\cite{shen10}, the accretion induced collapse of a white dwarf to a neutron star\cite{dessart06,darbha10}, the merger of two neutron stars (i.e. kilonova) \cite{Abbott17,Villar17} (see Figure~\ref{fig:absmag}), or the core collapse of massive stars if little ejecta is produced\cite{tauris15,moriya10}. While radioactive  models of these scenarios can reproduce the timescales observed for the KSN2015K light curve, they fail to reproduce its peak brightness\cite{shen10,kasen15,moriya16,darbha10,piro14} on rather general physical grounds. The peak luminosity of a radioactive supernova is approximately (to within a factor of $\sim2$) given by the instantaneous rate of heating by decay. The heating rate for radioactive isotopes with half-lives in excess of a few days (such as $^{56}$Ni at $\sim3\times10^{10}~{\rm ergs~s^{-1}~g^{-1}}$) requires a radioactive mass of $\sim 0.1~M_\odot$ to power the peak of KSN2015K. This conflicts with the $\sim 10^{-2}~M_\odot$ limit on the {\it total} ejecta mass inferred from the light curve risetime. This tension is not likely resolved by arguing for an anomalously low opacity in KSN2015K, since the luminosity and hence thermal state are similar to that of ordinary SNe. Figure~\ref{fig:risetime} quantifies the allowed range of radioactive powered light curves and shows that such a source can be ruled out for  KSN2015K. 

An alternative possible power source for supernova light curves is energy deposition from a central engine, such as a rotating magnetized neutron star\cite{maeda07,kasen10} (a magnetar) or an accreting black hole\cite{dexter13}. Such compact objects may be formed in the core collapse of a rotating massive star, and have been suggested to power the most luminous supernovae. The ejecta mass constraints above apply to central engine heating, but the peak luminosity can be substantially greater than is possible with radioactivity. However, explaining KSN2015K with a central engine implies extreme or fine-tuned parameters. A magnetar with rotational energy $E_m$ and spindown time $t_m$ produces a peak light curve luminosity of approximately\cite{kasen_bildsten_2010} $L\sim~E_m~t_m/t_{\rm diff}^2$. For an aligned force free wind\cite{metzger14}, the quantity $E_m~t_m$ is independent of the magnetar spin period and depends only on the surface equatorial dipole magnetic field, $B$,  as $E_m t_m \approx 6 \times 10^{84} B^{-2}$. The properties of KSN2015K ($L\approx10^{43}~{\rm ergs~s^{-1}}$ and $t_{\rm{diff}}\approx2$~days) then suggest an extreme field of order $B \sim 5 \times 10^{15}$~Gauss. A magnetar model could be constructed to fit the light curve of KSN2015K, but it would require invoking both an exceptionally strong magnetar and an unusually small ejecta mass.

For a black hole model, the small ejecta mass of KSN2015K would indicate a nearly failed supernova where all but $\lesssim 1\%$ of the star remained bound to the black hole. The  power from fallback accretion can be estimated\cite{chevalier89} as  $P=\epsilon~M_{\rm{fb}}/t_{\rm{fb}}~(t/t_{\rm{fb}})^{-5/3}$, where $M_{\rm{fb}}$ is the fallback mass, $t_{\rm{fb}}$ the fallback time, and $\epsilon$ the accretion efficiency. For $M_{\rm{fb}}\approx\Msun$ and adopting a relatively short fallback time $t_{\rm{fb}}\sim1$~hour (characteristic of compact stripped star with $R\sim~R_\odot$) the  accretion power will far exceed the luminosity of KSN2015K at $t=2$~days unless the efficiency is $\epsilon\sim10^{-5}$, which is much less than the characteristic value $\epsilon\sim0.1$. To reconcile the difference would require fine tuning the fallback dynamics and/or accretion disk formation such that only a tiny fraction of the infalling material was tapped to power the light curve.
 
Long gamma-ray bursts (GRBs) result from the core-collapse of very massive stars\cite{stanek03} that drive collimated relativistic jets. When a jet is viewed off axis, no gamma-rays are seen, but the shocked circumstellar gas may be visible as an ``orphan afterglow''. The light curve of KSN2015K is a good match to orphan afterglow models (see the Supporting Material). However, GRBs are very rare compared to SNe, so the chance of having found a GRB afterglow during the K2 mission is exceedingly small (see Supporting Materials).

A final class of models for KSN2015K suggests that the transient is powered by energy deposited by a hydrodynamical shock, either the shock of the supernova explosion itself or one occurring post-explosion due to the interaction of the stellar ejecta with the circumstellar medium (CSM)\cite{chevalier11,balberg11,ofek10,ginzburg14,kleiser14}. An explosion shock carries energy to the outer layers of the star and eventually vents in a shock breakout event at a radius $R$ where the optical depth $\tau$ is low enough that the radiative diffusion timescale, $t_d\approx\tau~R/c$, becomes comparable to the dynamical time, $R/v_s$, where $v_s$ is the shock velocity. This occurs at an optical depth $\tau\approx~c/v_s\approx30$ for a shock velocity $v_s=10^4~{\rm{km}~s^{-1}}$.

To explain the rapid rise of KSN2015K as a shock breakout event requires that the diffusion time from the shock $t_d\approx30~R/c$ be of order 2~days, which implies $R\approx2\times10^{14}$~cm. This is larger than typical radii of red supergiant supernova progenitor stars\cite{dessart13}. The effective radii of red supergiants could be increased just prior to explosion by envelope inflation or enhanced mass loss through winds. However, if the progenitor had been a supergiant with a wind, the explosion would have resulted in a long lasting light curve similar to a Type~IIP at later times\cite{moriya17} (i.e.,  $L\approx10^{42}~{\rm{ergs~s}^{-1}}$ at $t\approx50$~days),  which is inconsistent with the rapid dimming of KSN2015K. We therefore conclude that the progenitor was more compact (e.g., a helium or carbon/oxygen star) with radius $\approx 10^{11}$~cm and interacted with a dense and extended CSM at radius of several times $10^{14}$~cm. Shock breakout thus occurs in the extended CSM shell\cite{chevalier11}.

We can make an order of magnitude estimate of the minimum mass loss rate required to explain the KSN2015K light curve.
Assuming constant density, the CSM mass required to produce $\tau\approx~c/v$ is  $M\approx4\pi~R^2~c~\kappa~v_s \approx    10^{-2}~M_\odot$ for $\kappa=0.34~{\rm{cm}^2~\rm{g}^{-1}}$. This CSM must have been lost within a time $t_{\rm{csm}}\approx R/v_{\rm{csm}}$ before explosion, where $v_{\rm{csm}}$ is the CSM velocity. For $v_{\rm{csm}}=10~{\rm{km~s}^{-1}}$ (typical of a red-giant wind) we have $t_{\rm{csm}}\approx6$~years and an effective mass loss rate of $\dot{M}\approx2\times10^{-3}~M_\odot~{\rm{yr}}^{-1}$.  For the more likely case of a stripped envelope progenitor, the characteristic escape velocity is $v_{\rm{csm}} = 1000~{\rm{km~s}^{-1}}$ which implies a  mass loss episode with $\dot{M} \approx 2 \times 10^{-1}~M_\odot~{\rm yr}^{-1}$ occurring $t_{\rm{csm}} \approx 20$~days before the explosion. Such mass loss rates are much greater than typical winds from massive stars, but could be produced in episodic mass loss outbursts.

To test whether the shock breakout in CSM can explain the light curve of KSN2015K, we ran numerical radiation-hydrodynamical simulations of a supernova running into a circumstellar shell (see SM). Figure~\ref{fig:lightcurve} shows that for a model with CSM masses and radii roughly in the range estimated above, the venting of the post-shock  energy at breakout can explain KSN2015K's very rapid rise to a luminous peak.
The post maximum luminosity is due to the diffusion of shock deposited energy from deeper layers. At later times ($t\gtrsim10$~days) the decline of the KSN2015k light curve becomes shallower and it is possible that radioactive $^{56}$Ni decay contributes to the luminosity. 
The numerical calculation suggests a somewhat higher CSM mass ($\approx 0.15~M_\odot$) than the simple minimum mass analytic estimates above, although this number can depend upon specific details of opacity and composition.
As the shape and brightness of the model light curves are sensitive to the CSM and ejecta parameters (Supplementary Figure~1) the full coverage high sampling of the KSN2015K light curve provides strong constraints on the conditions of shock breakout in a dense circumstellar medium.


Fast transients are difficult to discover and follow-up, and sufficient numbers have been discovered only in recent years due to surveys with improved cadence and depth like Pan-STARRS1 (PS1) and Palomar Transient Factory (PTF). One of the earliest fast-transients identified was SN2002bj, which was initially postulated to be a ``.1a'' event\cite{poznanski10}, but its high luminosity makes this unlikely. The spectrum of SN2002bj was similar to a SNIa except for a prominent Helium line suggesting that it might be a stripped core-collapse event with a Helium envelope. The very bright SN2015U rose in less than 10 days and its time above half maximum was $t_{1/2}=12$ days. It showed narrow Helium features\cite{shivvers16}, implying that interaction with a hydrogen-poor CSM does occur in rapidly evolving events. Similarly, SN2010X had a rise of less than 10 days and $t_{1/2}=15$ days, but was four times fainter than SN2002bj\cite{poznanski10,kasliwal10}. The rapid evolution and lower luminosity means SN2010X could be powered by radioactive decay of thermonuclear products. The blue and fast transient iPTF~16ASU\cite{Whitesides17} has a comparable color and rise-time than KSN2015K, but it is significantly brighter and the overall event duration is also longer by at least a factor of two. In Figure~\ref{fig:absmag}, we compare the light curve of KSN2015K with SN2002bj and SN2015U.

The largest sample of fast transients\cite{drout14}, discovered in the PS1 survey, has rise time upper-limits of 3 to 5 days and peak luminosities similar to KSN2015K (see Figure~\ref{fig:risetime} for a comparison of rise times and absolute magnitudes). These PS1 transients also show very blue colors with typical $g-r = -0.2$ mag near maximum light and only a slow reddening afterwards. Thus, the PS1 transients are very similar to KSN2015K in all their photometric properties. Using the \cite{drout14} FELT rate from PS1 (see Supporting~Material), we expect to find a small number of FELTS in K2. Fast transients from the Supernova Legacy Survey\cite{arcavi16}, the Palomar Transient Factory\cite{arcavi16}, and the Subaru telescope\cite{Tanaka16}, are brighter, have significantly longer rise times and/or longer event durations, and are therefore likely to be different to FELTs.


We find that KSN2015K and the fast transients from the PS1 sample are most consistent with the shock-breakout into a dense circumstellar shell. It reproduces the significant characteristics of FELTs (fast, bright, blue) without much fine tuning. Even though models with a central engine can fit the light curve of KSN2015K and other FELTs, it requires an unlikely confluence of rare occurrences, and therefore is less likely. All of the other models of the power source of these events cannot explain at least one of their main properties.

\clearpage

\clearpage

\putbib
\end{bibunit}

\begin{addendum}
 \item This work is partially supported by NASA K2 cycle 4 Grant NNH15ZDA001N and cycle 5 Grant NNX17AI64G. We acknowledge support from
the Australian Research Council Centre of Excellence for All-sky Astrophysics (CAASTRO), through project number CE110001020.
 \item[Competing Interests] The authors declare that they have no
competing financial interests.
 \item[Correspondence] Correspondence and requests for materials
should be addressed to A. Rest~(email: arest@stsci.edu).
 \item[Author Contributions] AR, PMG, BT, and DK contributed to the scientific analysis. DK compared the data to theoretical models. EJS discovered the KSN2015K event and he, RPO, and RM reduced the K2 light curve data. AZ, GS, DJ, and RCS took and reduced the DECam data. SM and BT obtained and reduced the spectra. FF and VAV contributed the light curve fitting. All authors contributed to the scientific text.
 \end{addendum}
 
\clearpage

\begin{bibunit}

\section*{Methods}

\subsection{Additional Photometry}

DECam on the CTIO 4-meter observed the transient on BJD=2457236.523, only 1.4 days after maximum light. Using template subtraction, an SDSS-i band magnitude of 19.55$\pm 0.02$ was measured for this observation at R.A.=13:31:51.64 and Dec=-10:44:09.48. Images taken 12 and 16 days later finds the transient has faded below the 5$\sigma$ detection limits of 22.41 mag and 22.87 respectively.

SKYMAPPER observed the transient on BJD=2457242.906 (7.8 days after maximum) and obtained magnitudes of 20.01$\pm 0.12$ and 20.18$\pm 0.17$ in SDSS-g and SDSS-r bands respectively. The $g-r$ color is $-0.17\pm 0.20$, which is quite blue despite having faded to half its peak brightness.

We can compare the DECam i-band magnitude to the K2 brightness obtained at the same time to approximate a SDSS $r-i$ color. The K2 bandpass can be approximated as $K_p = 0.3g+0.7r$ which is approximately $r$ when $g-r\approx 0$. So the K2 magnitude at the time of the DECam observation was r=19.40$\pm 0.05$ resulting in a color of $r-i = -0.15\pm 0.05$ mag.

\subsection{Rise Time}

Time of maximum light is 2402.09 = BJD -- 2457235.1$\pm 0.1$ estimated using 3rd order fit around peak. The earliest 3$\sigma$ detection is $-2.02\pm 0.02$ rest frame days from max light. The rise in flux, $F$, is well fit by a quadratic polynomial $$ F(t<0)/F_{peak} = -0.3550\; t^2-0.100\; t+1.00 $$ where $t$ is the age in days (a negative number before maximum). The fit reaches zero flux at $t=-1.83$ days. The quadratic model fails to fit the earliest significant detections before $t< -1.7$ days, so we add a ``toe'' rising like $t^2$ to join with the quadratic and allow the time of explosion to be a free parameter. The result suggests the explosion occurred at $-2.2\pm 0.1$ days before maximum light.

\subsection{Shock Breakout in Circumstellar Material (CSM) Model}
In order to  model the CSM shock-powered scenario, we carry out one-dimensional radiation hydrodynamical simulations in spherical symmetry using CASTRO \cite{almgren10,zhang11}.  We solve the equations of radiation hydrodynamics with a gray flux-limited non-equilibrium diffusion approximation. We model the supernova ejecta as a homologously expanding broken power-law\cite{kasen16}, characterized by an ejecta mass $M_{\rm ej}$, outer ejecta velocity $V_{\rm ej}$, and outer radius $R_{\rm ej}$, with a temperature $T_{\rm ej}=10^4$ K. The CSM is modeled as a constant-density shell with mass $M_{\rm csm}$ at radius $R_{\rm csm}$ and thickness $\Delta R_{\rm csm}$. The shell is initially ``cold'' with $T_{\rm csm}=10^3$ K and moving at $V_{\rm csm}=100$ km s$^{-1}$.  We assume the opacity is dominated by electron scattering, where $\kappa_{\rm es}=0.4$ cm$^2$ g$^{-1}$ for ionized hydrogen. Recombination effects are modeled by an opacity drop-off at the hydrogen recombination temperature $T_{\rm rec}\approx 5000$ K.

The best-fit CSM model of KSN2015K is shown in Supplementary Figure~1. The rise time is set by the shell thickness, $t_{\rm rise}\sim\Delta R/V_{\rm ej}\sim 2$ days and required a thin shell of $\Delta R_{\rm csm}/R_{\rm csm}\approx 0.25$ to capture the rapid rise of KSN2015K. Overall, the thin-shell CSM interaction model provides a reasonably good fit to the KSN2015K data, matching well the rise, peak, and decline of the light curve. A more systematic exploration of FELTs due to CSM shock interaction will be forthcoming (Khatami et al. in prep).

While the inferred parameters don't provide any constraint on the exact mass-loss mechanism to produce a thin dense shell, the best-fit model gives  reasonable ejecta parameters and CSM mass and radius. If we assume the mass loss is steady over a finite interval with a constant velocity of $V_{\rm csm}\sim 1000$ km s$^{-1}$, the CSM is ejected roughly $t_{\rm csm}\sim R_{\rm csm}/V_{\rm csm}\sim 0.13$ years prior to the supernova, and occurred over a time $\Delta t_{\rm csm}\sim \Delta R_{\rm csm}/V_{\rm csm}\sim 11$ days.

\subsection{Orphan Afterglow Model}

Long gamma-ray bursts (GRB) are the result of the core-collapse in massive stars \cite{stanek03} that eject highly collimated, relativistic jets \cite{stanek99} into the circumstellar environment of the progenitor. The jet generates a shock into the circumstellar gas that emits radiation at wavelengths between the x-ray and radio domains. The high Lorentz factor, $\gamma$, of the jet means that the shock emission is beamed. The optical light curves of GRB afterglows are generally decaying power laws with an initial index of $\alpha_1=1.2\pm 0.5$. The beaming angle increases as the shock slows and when the Lorentz factor becomes comparable to the inverse jet opening angle ($\gamma \approx \theta_{jet}^{-1}$) an on-axis observer can see the entire shock. At this point the power law decay slope increases to $\alpha_2=2.8\pm 0.3$ \cite{rhoads99}.

If we set the observed time of first light (BJD=2457232.70) to be $t=0$, then after maximum the KSN2015K light curve behaves like a broken power law (Supplementary Figure~2). The initial decay index is a shallow 0.5 with a sharp break at about 8 rest frame days where the decay index becomes 2.8. Unlike a GRB afterglow viewed near the jet axis, the light curve brightens over the first 2 days. For an orphan afterglow, the viewing angle, $\theta_{obs}$, is greater than the jet opening angle, $\theta_{jet}$, and no gamma-ray burst is seen, but afterglow emission may become detectable as $\gamma$ declines and the beaming angle increases \cite{granot02,totani02}. Thus, an orphan afterglow will rise quickly, reach a peak and then fade like a power law or broken power law. The Totani et al. model does an excellent job of matching the width of the transient's light curve with power law indices that are typical of decaying GRB afterglows ((Supplementary Figure~2). The model predicts a rise that goes as $t^8$, but KSN2015K increases in brightness more slowly. A better fit to the early rise can be accomplished by shifting the assumed time of the burst by about 6 hours before the transients's first light. Overall, an orphan afterglow light curve is a good match to KSN2015K. However, the ratio of GRB to core-collapse supernovae is of order 10$^4$, making it very unlikely that KSN2015K is an orphan afterglow.

GRBs are very rare compared to SNe, e.g. there is only 1 long GRB for every 1000 SNIb/c\cite{grieco12}. Further, SNIb/c constitute $\sim$10\%\ of all core-collapse supernovae in the local universe \cite{prieto08}. By definition, orphan afterglows are viewed offaxis at an angle greater than the jet opening angle: $R=\theta_{obs}/\theta_{jet} >1$. At a fixed luminosity, the orphan rate should be higher than the GRB rate by $\sim R^2$. However, the peak brightness of orphan afterglows decreases quickly with viewing angle, suppressing the chances of seeing an orphan event far off axis. For our GRB model of KSN2015K, we find $R\approx 2$. Increasing $R$ to 3 would suppress its brightness by a factor of ten\cite{granot02} making it undetectable in the K2 data. The rate of orphan afterglows with the properties of KSN2015K are, at most, a factor of about 10 higher than the rate of long GRB. Since there has only been a handful of core-collapse events detected by Kepler/K2, the chances of finding a GRB afterglow during the K2 mission is exceedingly small.

\subsection{FELT rates}

We estimate how many FELTs we should have expected to find in the K2 Campaigns up to Campaign 15. First, we integrate the galaxy stellar mass functions\cite{tomczak14} and integrate them back to a redshift of zero, estimating the local stellar mass density to be between
\begin{equation}
\rho_M(z=0) = 1.4 - 2.5 \times 10^8 \Msun \mathrm{Mpc}^{-3}
\end{equation}
Based on the PS1 sample\cite{drout14}, FELTs have a volumetric rate of 
\begin{equation}
R_{FELT} = 4800-8000\ \mathrm{events}\ \mathrm{yr}^{-1} \mathrm{Gpc}^{-3}
\end{equation}
We can combine these two measures to calculate the SN rate per unit mass\cite{Mannucci05}:
\begin{equation}
SNuM = \frac{R_{FELT}}{\rho_M(z=0)}
\end{equation}
We find the upper and lower limits of $SNuM$ to be between 0.02 and 0.06, where $SNuM$ has the units of (100 yr)$^{-1}$ $(10^{10}\Msun)^{-1}$. A given galaxy was observed for about 0.2 yr in a K2 campaign, with about 44,000 galaxies observed by Kepler in total until K2 Campaign 15. Assuming that the typical mass of these galaxies is $10^{10}\Msun$, we then estimate that we should have found between 1 and 5 FELTs up to Campaign 15.

\subsection{Host Galaxy Properties}

The host galaxy of KSN2015K was cataloged by the 2MASS survey as 2MASX-J13315109-1044061 with an infrared brightness of $K=13.57$ mag. For a luminosity distance of 410~Mpc ($H_0=70$ km~s$^{-1}$ Mpc$^{-1}$) and a $k$-correction of $-0.23$ mag, the infrared luminosity of the host is $M_K = -24.26$ mag. The luminosity function of 2MASS late-type galaxies derived by \cite{kochanek01} has a break at $M_{K*}= -23.75\pm 0.06$ ($H_0=70$ km~s$^{-1}$ Mpc$^{-1}$), showing that the host is a fairly luminous and massive galaxy. Such galaxies tend to have a gas-phase metal abundance close to solar.

Images of the host galaxy were obtained at the Keck telescope in 2016 June, after the transient had faded in $V$ and $I$ filters. The host was also imaged with the Large Binocular Camera (LBC) using the Large Binocular Telescope (LBT) through SDSS $g$, $r$, $i$, and $z$ filters, plus a Bessel-$U$ filter. A pseudo-color image was created from the $g$, $V$ and $I$ filters and is shown in the top panel of Supplementary Figure~3. The galaxy is clearly an inclined spiral with a blue color implying significant star formation. The transient was seen projected on a spiral arm 14.0~kpc from the host center.

A spectrum of the host was obtained with the Wide Field Spectrograph (WiFeS) on the ANU 2.3-m telescope (see bottom panel of Supplementary Figure~3). The spectrum covers a wavelength range from 360~nm to 800~nm with a resolution of 3000. The four spaxels nearest the location of the transient were combined to make a final spectrum covering one square arcsecond on the sky, which at the host redshift, corresponds to 1.7 $kpc^2$. Therefore, we have a relatively local spectrum of the host environment.

The spectrum shows a flat continuum with weak absorption features of Balmer series H$\beta$, Ca~II and Na~I. Strong emission lines of Balmer series H$\alpha$, [NII], [OIII], [OII], and [SII] are seen. The lines are shifted by $z=0.090\pm 0.002$ and we conclude that this is the redshift of the transient. This location in the host appears to contain a wide range of stellar ages. The Na~I absorption suggests an old stellar population while the H$\beta$ absorption likely comes from stars with an age of about 0.5 Gyr. The strong H$\alpha$ emission shows that the formation of new stars is on-going. The [OIII]/H$\beta$ ratio is consistent with solar metallicity as expected for a massive galaxy.

\putbib
\end{bibunit}

\setcounter{figure}{0}    
\renewcommand{\figurename}{Supplementary Figure} 

\begin{figure}
\begin{center}
\includegraphics[scale=0.32]{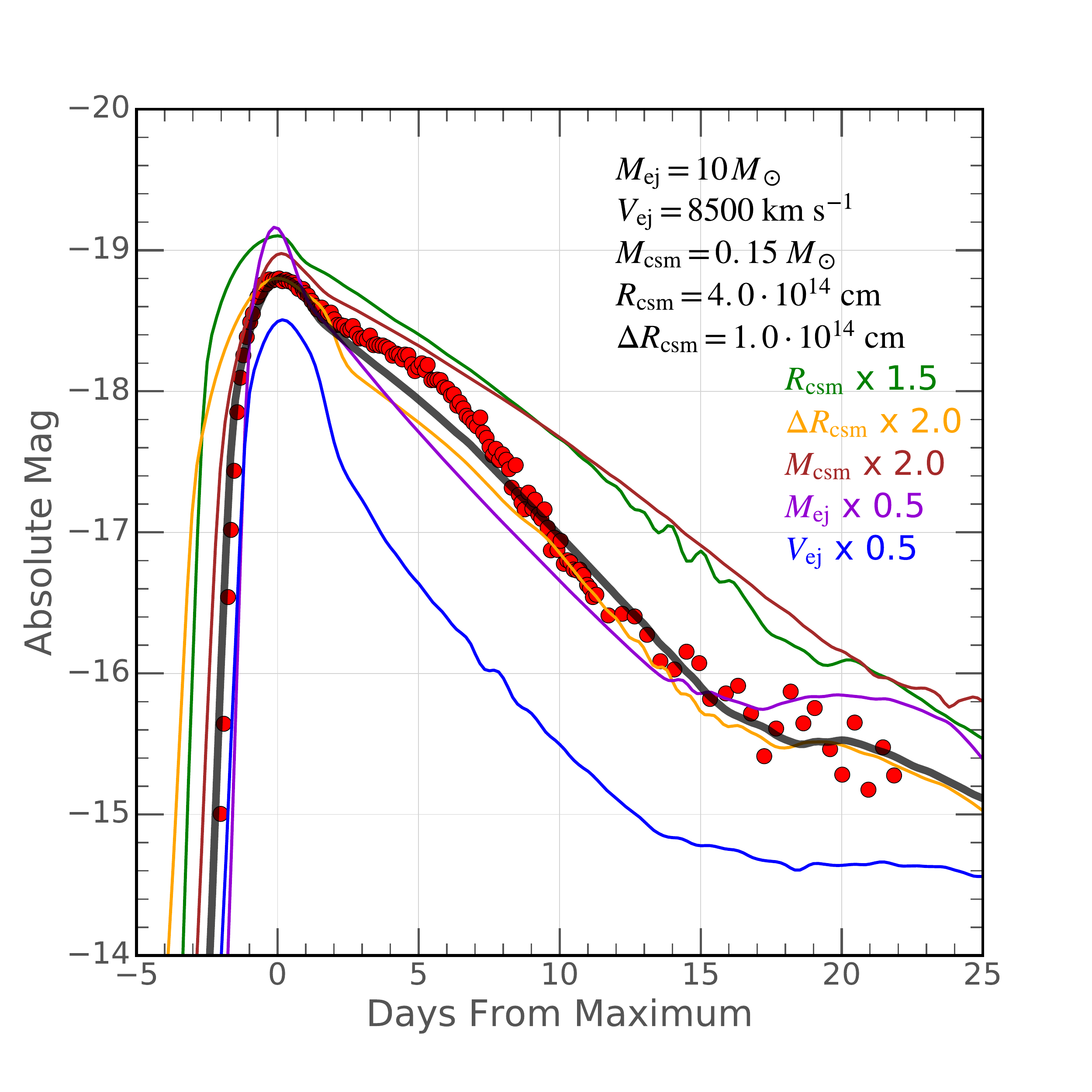}
\includegraphics[scale=0.32]{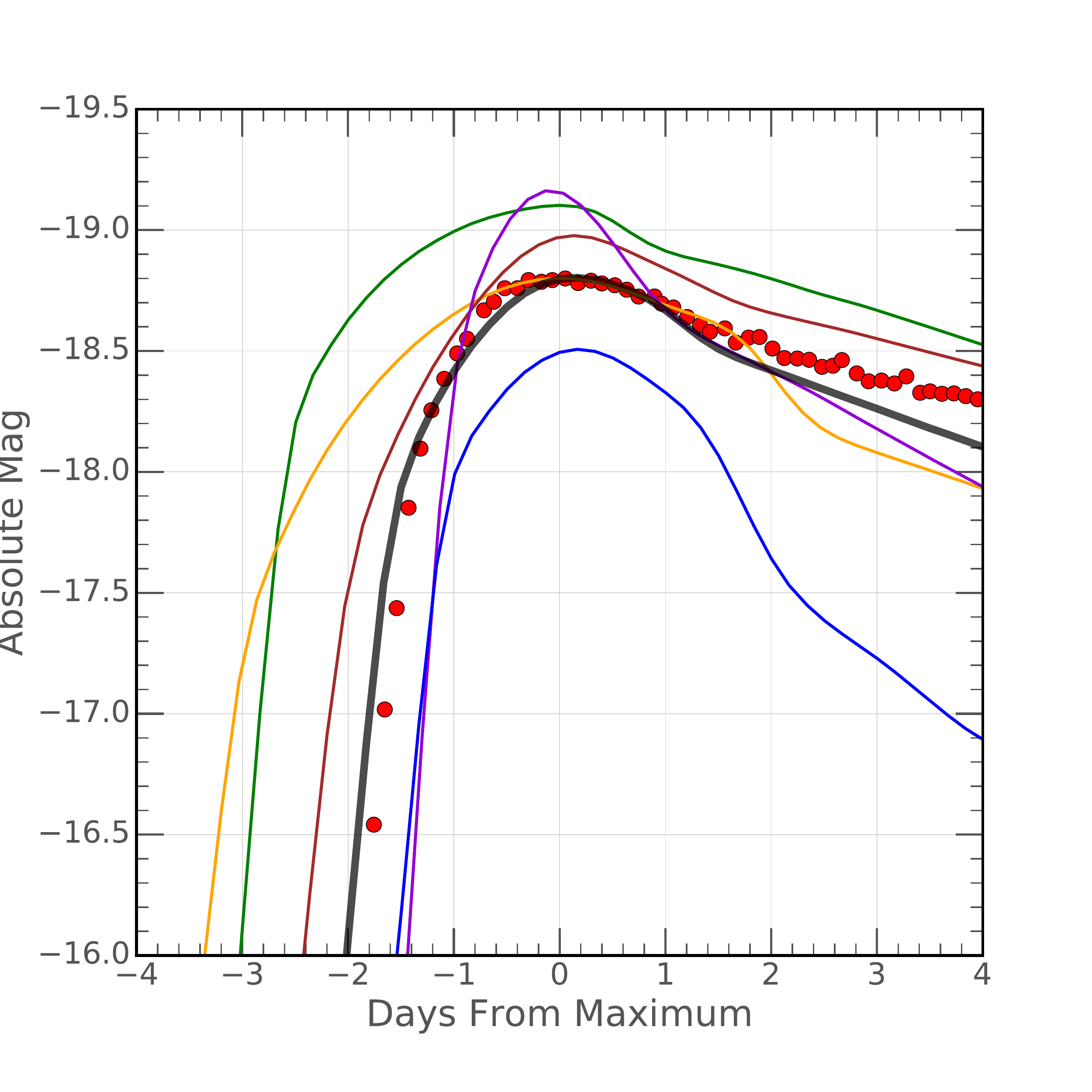}
\caption{{\bf CSM interaction models.} {\it Left:}  The KSN2015K data (red points) compared to numerical radiation hydrodynamics simulations of the CSM interaction models (lines).  The best-fit model (black line, parameters shown inset) is able to capture the fast rise and peak magnitude of KSN2015K as well as the rapid decline due to cooling of the shock-heated CSM and ejecta. {\it Right:} The CSM interaction model around peak.}
\label{fig:csm}
\end{center}
\end{figure}

\begin{figure}
\begin{center}
\includegraphics[scale=0.60]{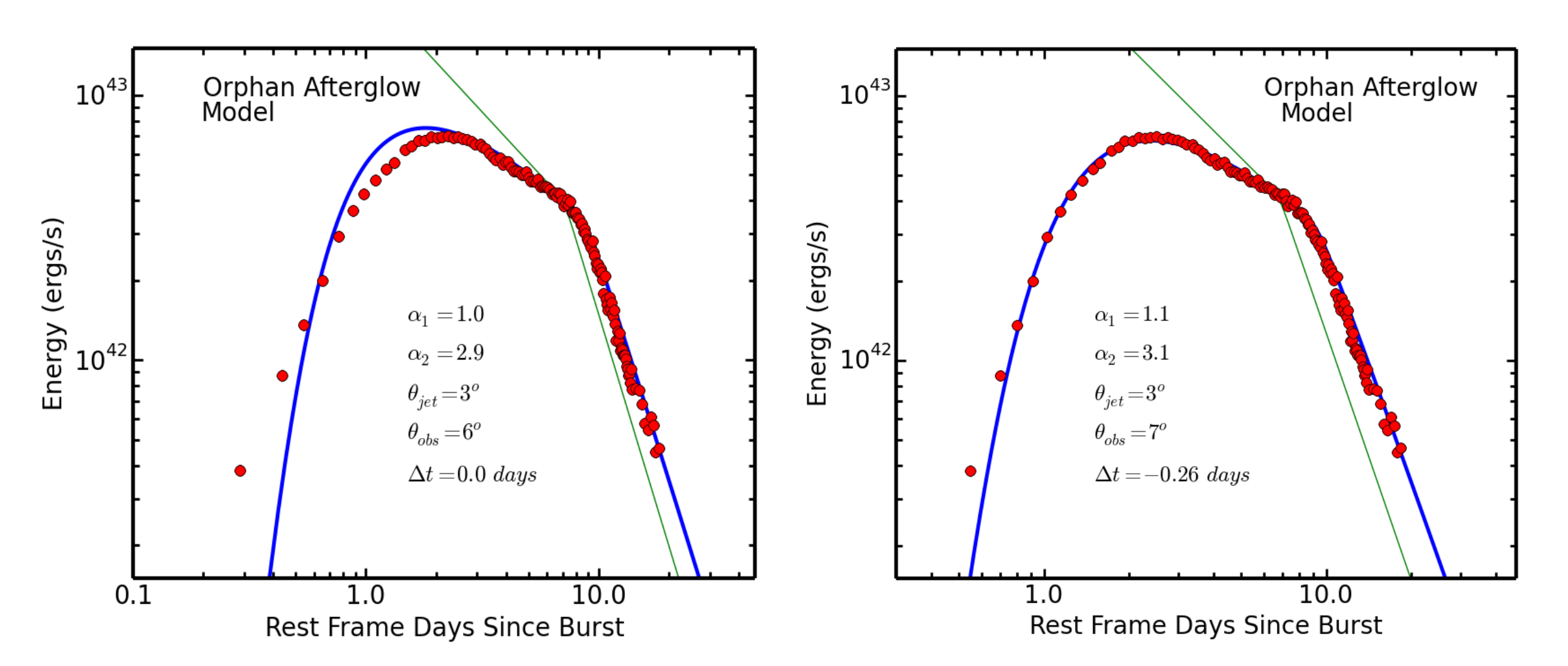}
\caption{{\bf Orphan afterglow models.} The KSN2015K light curve (red points) compared to orphan afterglow models (blue lines). Observing a GRB off the jet axis at an angle larger than the jet opening angle
means the gamma-rays are not seen, but the afterglow may become visible as the shock slows and the relativistic beaming broadens. The green lines show the model light curve if the GRB were viewed directly along the jet axis. {\it Left:} For the first model, the time of the GRB is assumed to be at the
moment of first light ($\Delta t=0$). However the orphan afterglow model rises too quickly to match
the observed light curve. {\it Right:} In the second model, the time of the GRB is assumed to be
about 6 hours before first light ($\Delta t=-0.26$ days). This has only a minor effect on the model itself, but forces the light curve rise more steeply and better match the model.
} 
\label{fig:orphan}
\end{center}
\end{figure}

\begin{figure}
\begin{center}
\includegraphics[scale=0.615]{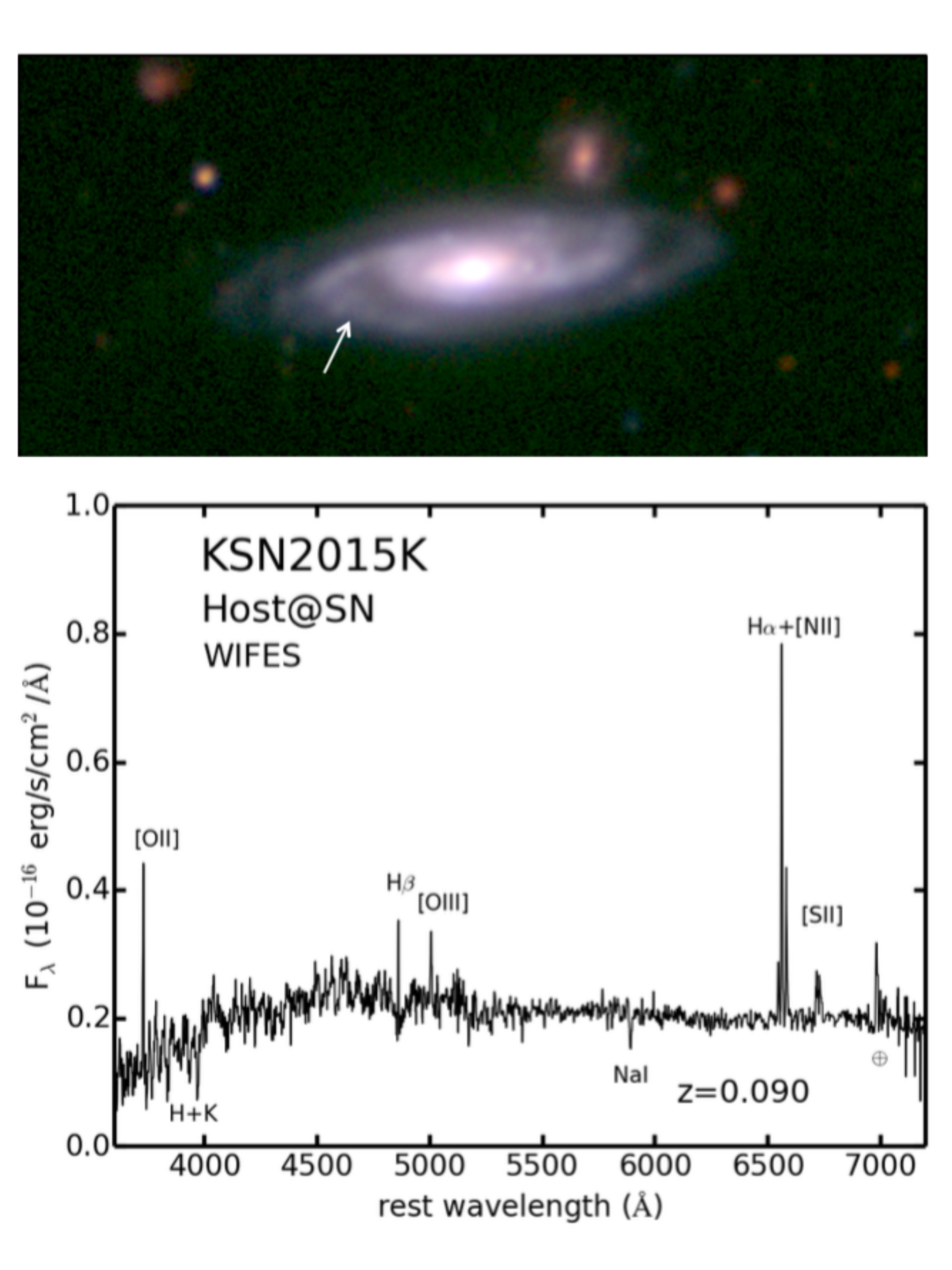}
\caption{{\bf Host galaxy 2MASX-J13315109-1044061.} {\it Top:} Pseudo-color image of the host galaxy of KSN2015K made using $V$ and $I$ band images from Keck and SDSS-$g$ band from the LBT. The arrow points to the position of the transient. {\it Bottom:} The WiFeS spectrum of the host galaxy.  The spectrum is the sum of four spaxels covering one square arcsecond at the location of the transient. The large H$\alpha$ equivalent width suggests on-going star-formation. The absorption at H$\beta$ implies a significant post-starburst population.
} 
\label{fig:host_color}
\end{center}
\end{figure}

\end{document}